\documentclass[prl,twocolumn,showpacs]{revtex4}
\usepackage{amsmath}
\usepackage[dvips]{graphicx}
\usepackage{amssymb}

\begin{document}
\title{Formation of Quantum-Degenerate Sodium Molecules}
\author{K. Xu, T. Mukaiyama, J.R. Abo-Shaeer, J.K. Chin, D.E. Miller, and W. Ketterle}

\affiliation{Department of Physics, MIT-Harvard Center for
Ultracold Atoms, and Research Laboratory of Electronics, MIT,
Cambridge, MA 02139}
\date{\today}

\begin{abstract}
Ultra-cold sodium molecules were produced from an atomic
Bose-Einstein condensate by ramping an applied magnetic field
across a Feshbach resonance. More than $10^5$ molecules were
generated with a conversion efficiency of $\sim$4\%. Using laser
light resonant with an atomic transition, the remaining atoms
could be selectively removed, preventing fast collisional
relaxation of the molecules.  Time-of-flight analysis of the pure
molecular sample yielded an instantaneous phase-space density
greater than 20.

\end{abstract}


\maketitle

Atomic Bose-Einstein condensates (BEC) provide a new window into
macroscopic quantum phenomena \cite{BEC99var}. A molecular
condensate could lead to a host of new scientific explorations.
These include quantum gases with anisotropic dipolar interactions,
tests of fundamental symmetries such as the search for a permanent
electric dipole moment, study of rotational and vibrational energy
transfer processes, and coherent chemistry, where reactants and
products are in coherent quantum superposition states. So far, the
highly successful techniques for creating atomic BEC have not led
to success for molecules. Laser cooling is difficult due to the
complicated level structure of molecules \cite{bahns96}, and
evaporative cooling requires the preparation of a dense gas of
molecules, where elastic collisions dominate inelastic collisions.

Alternative techniques, such as buffer gas loading \cite{wein98}
and Stark deceleration \cite{bethlem99}, have been successful in
obtaining cold molecules. Yet these methods are still far from
achieving the requisite phase-space density for BEC. The
difficulty in cooling molecules directly can be circumvented by
creating ultracold molecules from quantum-degenerate atomic
samples. This requires molecule formation without release of
energy, which can be accomplished either by photoassociation
\cite{wyna00} or by ``tuning'' a molecular state via a Feshbach
resonance \cite{inou98} to be degenerate with the atomic state. A
Feshbach resonance occurs when an applied magnetic field Zeeman
shifts a molecular state to zero binding energy. By ramping an
external field across a Feshbach resonance from negative to
positive scattering length, translationally cold molecules in high
vibrational states can be created adiabatically \cite{mies00,
abeelen99, yurovsky99}.

\begin{figure}[tbp]
\includegraphics[width=85mm]{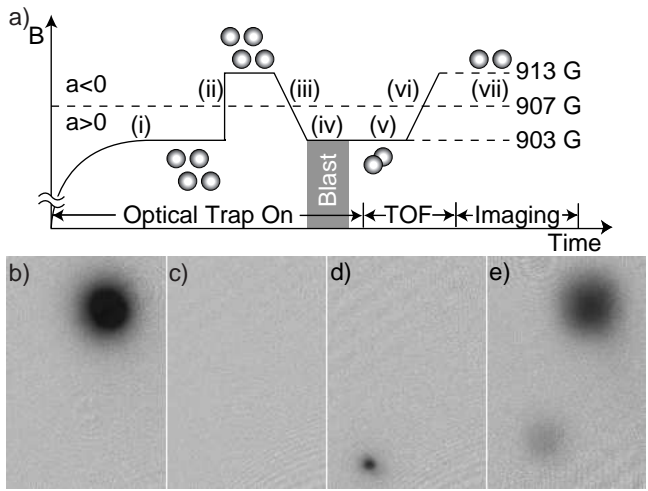} \caption{(a)
Experimental method for producing and detecting ultracold
molecules. (i) Bose condensed atoms in an optical dipole trap are
exposed to a magnetic field just below a Feshbach resonance.  (ii)
The field is quickly stepped through the resonance to minimize
atom loss.  (iii) The field is then swept back through the
resonance, creating an atom-molecule mixture. (iv) Unpaired atoms
are removed from the trap with resonant light, yielding a pure
molecular sample. (v) The trap is switched off, allowing the
molecules to expand ballistically. (vi) Finally, the magnetic
field is swept back across the resonance to reconvert the
molecules to atoms for imaging (vii). (b) Image of the
\emph{atomic} sample after ramping the field to produce molecules;
(c) after the resonant light pulse has removed all unpaired atoms;
(d) after the \emph{molecules} ($\sim10^5$) have been reconverted
to atoms. (b-c) were taken along the weak axis of the trap after
17 ms ballistic (time-of-flight -- TOF) expansion. (e) An image
showing both atomic (top) and molecular (bottom) clouds after 14
ms ballistic expansion, spatially separated by a magnetic field
gradient. With 4 ms field ramp-down time, some molecules survived
even without the blast pulse, but are much more heated. The field
of view of each image is 1.8~mm $\times$ 1.3~mm.}
\label{fig:Figure1}
\end{figure}

The first observation of a Feshbach resonance in ultracold atoms
showed a high rate of atom loss \cite{inou98, sten98stro}.
Theories accounted for this loss by assuming the formation of
ultracold molecules \cite{mies00, abeelen99, yurovsky03mole}.
These molecules were predicted to decay vibrationally in less than
$100$~$\mu$s due to a two-body rate coefficient of order
$10^{-10}$~cm$^3$/s. Because of this, no successful attempt was
made to detect a molecular signature until atom-molecule beats
were observed in $^{85}$Rb, lasting about $100$~$\mu$s
\cite{donley02}. Recent fermion experiments using magnetic field
sweeps have observed molecules with lifetimes approaching 1 sec
\cite{regal03, cubizolles03, jochim03, strecker03}.  Until now,
similar experiments with bosons have only been carried out during
ballistic expansion \cite{durr03, herbig03}. According to theory,
the decay of molecules composed of fermionic atoms is suppressed
by Pauli blocking \cite{petrov03}, whereas molecules composed of
bosons decay rapidly.  This could explain the low conversion
efficiency of about 5\% for bosons, compared to $>$50\% for
fermions, where more adiabatic field ramps are possible.

If highly degenerate atoms (both fermionic and bosonic) are
converted adiabatically to molecules, the molecules can be created
at a phase-space density exceeding 2.6, the critical value at
which a uniform, ideal Bose gas condenses \cite{huan87}. Previous
experiments \cite{herbig03, cubizolles03, BEC03spain} have
measured or estimated conditions close to or around this critical
phase-space density.

Here we report the production of trapped sodium molecules from an
atomic BEC.  The initial phase-space density of the molecular
sample was measured in excess of 20. High phase-space density
could only be achieved by rapidly removing residual atoms, before
atom-molecule collisions caused trap loss and heating. This was
accomplished by a new technique for preparing pure molecular
clouds, where light resonant with an atomic transition selectively
``blasted" unpaired atoms from the trap. In contrast to spatial
separation via a Stern-Gerlach method \cite{herbig03, durr03},
this technique can separate out the molecules faster and does not
require a large difference in the magnetic moments of the atoms
and molecules.

To generate the molecules, sodium condensates in the
$\left|F=1,m_F=-1\right\rangle$ state were prepared in an optical
dipole trap. The radial and axial trap frequencies of $\omega_r =
2\pi\times290$ Hz and $\omega_z = 2\pi\times2.2$ Hz, respectively,
gave Thomas-Fermi radii of $R_r = 5$~$\mu$m and $R_z =
650$~$\mu$m, and a peak density of $1.7\times10^{14}$~cm$^{-3}$
for 5 million atoms. An adiabatic radio frequency sweep was used
to transfer the atoms into the $\left|1,1\right\rangle$ state,
which has a 1~G wide Feshbach resonance at 907~G \cite{inou98,
abeelen99na}.

After 1 second equilibration in the optical trap, the molecules
were generated using the field ramping scheme illustrated in Fig.
\ref{fig:Figure1}a. An applied magnetic field was ramped in
$\sim$100~ms to 4~G below the 907~G Feshbach resonance. The field
was generated using a pair of large bias and small anti-bias
coils. Because molecules are only created when sweeping across the
resonance from negative to positive scattering length, the field
was stepped up to 913~G as quickly as possible ($\sim$1~$\mu$s) to
jump over the resonance with minimal atom loss. After allowing
2.5~ms for transient field fluctuation to damp out, the field was
ramped down in time $\tau_{Down}$. Due to atom-molecule coupling,
part of the atomic population was transferred into the molecular
state following the Landau-Zener avoided crossing. With the given
width of the resonance and the atomic density, we use a simple
Landau-Zener model to calculate a ramp speed of $\sim$$10^4$~G/s
to transfer roughly half the atoms to the molecular state
\cite{mies00, abeelen99, yurovsky03mole}. However, inelastic
collisions led to fast decay for both the atoms and the molecules
near the resonance. We found that a faster ramp speed of
$\sim$$10^5$~G/s (corresponding to $\tau_{Down}=50$~$\mu$s) gave
optimal results. The conversion efficiency of atoms to molecules
was $\sim$4\%. Slower ramp speeds resulted in a similar number of
molecules, but at higher temperature (see Fig.
\ref{fig:Figure1}e).

The blast pulse was applied along the radial axis of the trap to
minimize collisions between the escaping atoms and the molecules
at rest. A $20$~$\mu$s pulse of resonant light removed all atoms
from the optical trap, leaving behind a pure molecular sample (see
Fig. \ref{fig:Figure1}). At only 4~G below the Feshbach resonance,
the light was still close to resonance with molecular
photodissociation to low-velocity atoms, but the overlap matrix
element was sufficiently diminished to leave the molecules
unaffected.  After a variable hold time, the optical trap was
switched off and the molecules expanded ballistically for between
4 to 20~ms. The molecules were detected by converting them back to
atoms with field ramp-up in $\tau_{Up}=100$~$\mu$s at the end of
expansion. Varying $\tau_{Up}$ between $50$~$\mu$s and 4 ms did
not affect the recovered atom number, though shorter $\tau_{Up}$'s
recovered atoms with larger kinetic energy \cite{takashi}. Thus we
assume all molecules are converted back to atoms. A resonant
absorption image was taken after an additional $500$~$\mu$s, which
allowed the imaging field to settle. The rapid conversion of
molecules to atoms after a long expansion time ensured that the
absorption images accurately depicted the momentum distribution of
the \emph{molecular} cloud.

\begin{figure}[tbp]
\includegraphics[width=85mm]{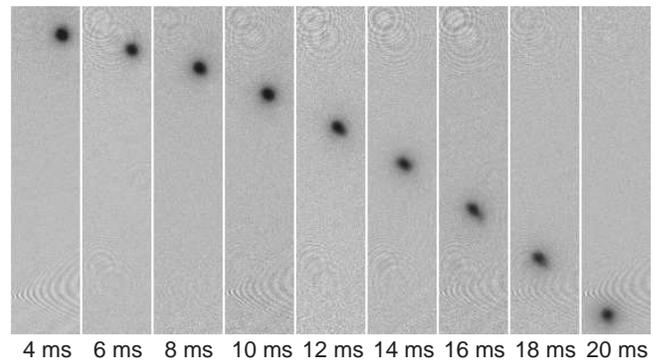} \caption{Ballistic
expansion of a pure molecular sample.  Absorption images of
molecular clouds (after reconversion to atoms) are shown for
increasing expansion time after switching off the optical trap.
The small expansion velocity corresponds to a temperature of
$\sim$30~nK, characteristic of high phase-space density. The
images are taken along the weak axis of the trap. The field of
view of each image is 3.0~mm $\times$ 0.7~mm.} \label{fig:Figure2}
\end{figure}

Atoms and molecules were separated during the ballistic expansion
by a Stern-Gerlach technique (Fig. \ref{fig:Figure1}e). Due to
trap imperfections, the large bias coils provided an additional
radial gradient of the axial field of $\sim$2.8~G/cm in the
vicinity of the condensate. This value was determined from the
trajectory of the falling atoms. Since the molecules have a
different magnetic moment, they separate from the atoms during the
ballistic expansion (Fig. \ref{fig:Figure1}e). From the separation
of the atomic and molecular clouds at different times, we
determined the difference between atomic and molecular magnetic
moments to be 3.2$\mu_B$ ($\mu_B$ is the Bohr magneton), in good
agreement with theory \cite{yurovsky03mole}.

For different ramp down times $\tau_{Down}$, the time-of-flight
images of the molecular cloud exhibit drastically different
momentum distribution. The coldest cloud was obtained with the
fastest ramp down time possible, $\tau_{Down}=50$~$\mu$s (Fig.
\ref{fig:Figure2}). A gaussian fit was used to determine the
molecular temperature $T_m$ and the phase-space density. Due to
the rapid ramp down, the molecules had no time to adjust to the
external trapping potential or any mean-field interactions.
Therefore, we assume the molecules were \emph{uniformly} created
with the Thomas-Fermi profile of the original atomic BEC. The peak
phase-space density is then given by
\begin{equation}
PSD_{peak}=\left({\frac{h}{\sqrt{2\pi k_B T_m
M_m}}}\right)^3\frac{N_m}{\frac{8\pi}{15}R_r^2 R_z} \label{eq1}
\end{equation}
where $h$ is the Planck constant, $k_B$ is the Boltzmann constant,
$M_m$ is the molecular mass, $N_m$ is the number of molecules. The
second factor in the equation is the peak density for a
Thomas-Fermi profile.

\begin{figure}[tbp]
\includegraphics[width=85mm]{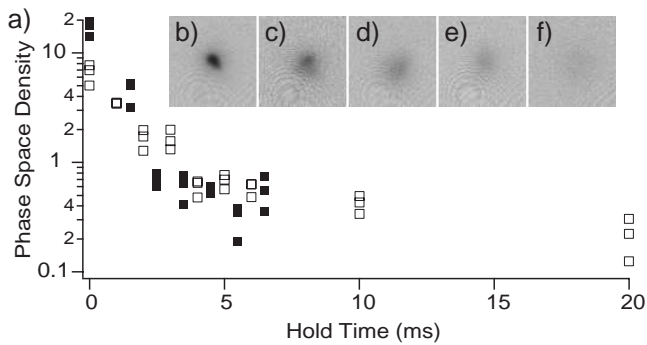}
\caption{Molecular phase-space density versus hold time. (a) The
phase-space densities of the trapped molecules were observed to
decrease significantly after a few milliseconds in the optical
trap. The open and solid squares are data from two separate runs
on different days. (b-c) are absorption images of the molecular
clouds after (b) 0 ms (c) 2 ms (d) 5 ms (e) 10 ms (f) 20 ms hold
time in the trap.  The field of view is 0.8~mm $\times$ 0.8~mm.}
\label{fig:Figure4}
\end{figure}

Fig. \ref{fig:Figure4}a shows the phase-space densities obtained
for different holding time in the optical trap. Phase-space
densities in excess of 20 were observed, much larger than the
critical value of 2.6.  This demonstrates that a
quantum-degenerate cloud of atoms can be transformed into a
quantum-degenerate molecular gas.

The high initial phase-space density decayed rapidly ($\sim$2~ms),
due to molecule loss and heating. For a pure molecular sample at a
peak density of $4\times10^{12}$~cm$^{-3}$, the molecule number
dropped by half in 5 ms and the apparent temperature doubled in 2
ms. Since the molecules are formed in a high vibrational state
with quantum number $v=14$, losses are most likely due to
vibrational relaxation. The high loss rate of the molecules is
consistent with theoretically predicted 2-body relaxation rate
coefficients of $10^{-10}$~cm$^3$/s \cite{yurovsky99, soldan02}.
Because the loss of molecules is faster at the high densities near
the bottom of the trap, it is accompanied by heating. This is in
contrast to evaporative cooling, where the losses occur at the top
of the trap.  Such anti-evaporative heating gives a time constant
four times slower than the observed heating rate. We therefore
believe that the rapid increase in the apparent temperature is due
to the inward motion of the molecular cloud (see below), and
possibly transfer of the vibrational energy of the molecules.

Our calculation of the phase-space density is conservative, since
almost all errors lead to an underestimation of the value. The
most critical quantity is the thermal velocity $v_{therm}=\sqrt{2
k_B T_m/M_m}$ obtained from the gaussian fit of the cloud, since
the phase-space density scales with the third power of
$v_{therm}$. We determined the velocity by simply dividing the
size of the cloud by the time-of-flight, without correcting for
imaging resolution and initial cloud size.

\begin{figure}[tbp]
\includegraphics[width=85mm]{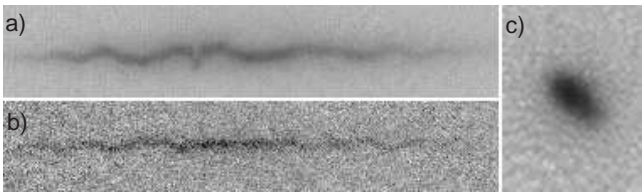}
\caption{Images of (a) atomic and (b) molecular clouds. These
absorption images were taken after 7~ms ballistic expansion and
show the axial extent of the clouds.  Radial excitations in the
optical trap resulting from the sudden switching of magnetic
fields are manifest as snake-like patterns. Such excitations blur
images (c) taken along the long axis of the trap (in 17~ms TOF),
leading to an underestimate of the phase-space density. The fields
of view are: (a) and (b) 0.6~mm $\times$ 3.2~mm (c) 0.6~mm
$\times$ 0.4~mm.} \label{fig:Figure3}
\end{figure}

Correcting for the imaging resolution of $10$~$\mu$m compared to
the typical cloud size of $50$~$\mu$m would increase the
phase-space density measurement by 6\%. In addition,  radial
excitation of the trapped cloud (shown in Fig. \ref{fig:Figure3})
contributed to the size of the cloud after the ballistic
expansion.  From the fits, the smaller of the two gaussian radii
was used to calculate $v_{therm}$, assuming that the larger size
was caused by radial excitations. Yet since the radial excitation
can occur in two orthogonal directions, we estimate that the
extracted thermal velocities were still overestimated by
$\sim$10\%. We also considered magnetic focusing of the cloud due
to residual field inhomogeneities. Because we use large coils
($\sim17$~cm in diameter and $\sim4$~cm away from the condensate)
to produce a homogeneous magnetic field, any residual radial
curvature due to radial fields is calculated to be
$\lesssim0.1$~G/cm$^2$. An upper bound for the radial curvature of
the axial fields was obtained from trap frequency measurements and
ballistic expansion measurements as $<1$~G/cm$^2$. This can only
reduce the size of the cloud by less than 2\% after a typical
ballistic expansion time of 17~ms.

We assume resonant absorption in determining the number of atoms.
Any systematic effect such as small detuning or saturation, would
lower both $N_m$ and the Thomas-Fermi volume (proportional to
$N^{\frac{3}{5}}$, where $N$ is the number of condensed atoms).
The net effect is an underestimate of the phase-space density. In
addition, because the molecular formation process is non-linear in
atomic density, the assumption of the atomic Thomas-Fermi volume
for molecules is likely an overestimate. Furthermore, in the
absence of strong mean-field repulsion (due to the much lower
molecular density), the molecular cloud would not sustain the
initial size of the atomic condensate (used in equation
(\ref{eq1})), and shrink to a smaller size within a few
milliseconds ($\sim$ radial trap period).  If we assume radial
thermal equilibrium while keeping the axial length fixed (as the
axial trap period is 500~ms), the phase-space density would be 2
to 4 times higher than is shown in Fig. \ref{fig:Figure4}. To sum
up, the extracted peak phase-space densities are underestimated by
$\gtrsim$30\%, and all other critical systematic effects would
raise the value even further.

When a molecular cloud with high phase-space density equilibrates
by elastic collisions, it should form a condensate. There is no
prediction for the scattering length of the molecules, which are
formed in the $\left|v=14,l=0\right\rangle$ state
\cite{moer95res}. Assuming a prototypical scattering length of $
100 a_0$ ($a_0$ is the Bohr radius), we estimate the elastic
collision rate between molecules to be 6~s$^{-1}$, which is
smaller than our loss rate. Thus, the so-called ratio of good and
bad collisions is smaller than one.

Recent work on molecules composed of fermionic lithium
\cite{cubizolles03,jochim03} and potassium \cite{regal03life}
atoms showed a dramatic increase in lifetime close to the Feshbach
resonance. Theoretically, the rate of vibrational relaxation
should decrease with the scattering length $a_s$ as $\varpropto
a_s^{-2.55}$ due to Pauli blocking \cite{petrov03}. In contrast,
for molecules composed of bosonic atoms, the rate should increase
proportionally to $a_s$ \cite{petrov03bos}. On the other hand, the
elastic collision rate is proportional to $a_s^2$, so for large
$a_s$ one would expect the ratio of good-to-bad collisions to
exceed one. However, if this condition is met at loss rates faster
than the trap frequency, the cloud can only establish local, not
global equilibrium.

Whether our molecular sample is a condensate depends on one's
definition of BEC. If phase-space density in excess of 2.6
(corresponding to a diagonal matrix element of the single-particle
density matrix larger than one) is sufficient, then one may regard
a short-lived atom-molecule superposition state \cite{donley02} as
a molecular BEC.  However, following this definition, a small
excited state admixture in an optically trapped BEC would qualify
as BEC of electronically excited atoms. If one asks for the
additional requirement of a pure molecular sample, we have
achieved that in this work. Another definition would require phase
coherence, which could again be observed even in short-lived
samples. Should one also require a lifetime of the degenerate
sample exceeding the collision time (to achieve local
equilibrium), the trap period (to achieve global equilibrium), or
the inverse mean-field energy (the typical dynamic timescale)? In
our opinion, BEC requires thermal equilibrium. High phase-space
density is necessary, but not sufficient.

In conclusion, we have created a quantum-degenerate gas of $10^5$
cold sodium molecules with a phase-space density $>$20. This was
achieved with a fast magnetic field sweep through a Feshbach
resonance, followed by quick removal of the remnant atoms with
resonant light. This purification was necessary to avoid heating
and decay of the molecules through inelastic collision processes.
These processes could also be avoided by loading the atomic BEC
into an optical lattice in the Mott-insulator phase with a filling
factor of two \cite{greiner02, jaksch02} which, after sweeping the
magnetic field through the Feshbach resonance, would result in a
long-lived sample of isolated molecules.

The authors would like to acknowledge M. Xue for experimental
assistance with the project.  We also thank A.E. Leanhardt and
M.W. Zwierlein for their critical reading of the manuscript. This
research is supported by NSF, ONR, ARO, NASA, and the David and
Lucile Packard Foundation.

\end{document}